\documentclass[reprint,aps,prl,superscriptaddress,nolongbibliography]{revtex4-2}
\pdfoutput=1
\usepackage{amsmath,amssymb}
\usepackage{graphicx}
\usepackage[T1]{fontenc} 
\usepackage{tikz-cd}
\usepackage{tabularx}
\usepackage{subfigure}
\newcolumntype{C}{>{\centering\arraybackslash}X}%
\newcommand{\uone}{\mathrm U(1)}
\DeclareMathOperator{\id}{id}

\DeclareMathOperator{\iTO}{iTO}

\begin{document}

\title{Computing classification of interacting fermionic symmetry-protected topological phases using topological invariants}

\author{Yunqing Ouyang}
\affiliation{State Key Laboratory of Surface Physics, Fudan University, Shanghai 200433, China}
\affiliation{Center for Field Theory and Particle Physics, Department of Physics, Fudan University, Shanghai 200433, China}
\author{Qing-Rui Wang}
\thanks{Present address: Yau Mathematical Sciences Center, Tsinghua University, Beijing 100084, China}
\affiliation{Department of Physics, Yale University, New Haven, CT 06511, USA}
\author{Zheng-Cheng Gu}
\email{zcgu@phy.cuhk.edu.hk}
\affiliation{Department of Physics, The Chinese University of Hong Kong, Shatin, New Territories, Hong Kong, China}
\author{Yang Qi}
\email{qiyang@fudan.edu.cn}
\affiliation{State Key Laboratory of Surface Physics, Fudan University, Shanghai 200433, China}
\affiliation{Center for Field Theory and Particle Physics, Department of Physics, Fudan University, Shanghai 200433, China}
\affiliation{Collaborative Innovation Center of Advanced Microstructures, Nanjing 210093, China}
\date{\today}

\begin{abstract}
  In recent years, great success has been achieved on the classification of symmetry-protected topological (SPT) phases for interacting fermion systems by using generalized cohomology theory. However, the explicit calculation of generalized cohomology theory is extremely hard due to the difficulty of computing obstruction functions.
  In this paper, based on the physical picture of topological invariants and mathematical techniques in homotopy algebra, we develop an algorithm to resolve this hard problem.
  It is well known that cochains in the cohomology of the symmetry group, which are used to enumerate the SPT phases, can be expressed equivalently in different linear bases, known as the resolutions.
  By expressing the cochains in a reduced resolution containing much fewer basis than the choice commonly used in previous studies, the computational cost is drastically reduced.
  In particular, it reduces the computational cost for infinite discrete symmetry groups, like the wallpaper groups and space groups, from infinity to finity.
  As examples, we compute the classification of two-dimensional interacting fermionic SPT phases, for all 17 wallpaper symmetry groups.
\end{abstract}

\maketitle
%

Topological phases~\cite{WenRMP2017,WenScience2019} are quantum states of matter beyond the Landau paradiam of classifying phases through spontaneous symmetry breaking.
Among them, the so called symmetry-breaking topological (SPT) phases~\cite{Gu2009,ChenCZX,ChenSPTScience,ChenSPTPRB} are distinguished from topologically trivial phases through symmetry-protected gapless edge states (or more generally, symmetry anomalies on the edge) and nontrivial responses to the insertion of symmetry fluxes, while the bulk is short-range entangled and lacks of fractionalized excitations.
Examples of SPT phases include topological insulators (TI)~\cite{Hasan2010,Qi2011}, topological superconductors (TSC), topological crystalline insulators (TCI)~\cite{Fu2011} and one-dimensional(1D) Haldane chain~\cite{Affleck1987}, which all have been realized in solid-state materials.

The classification of interacting fermion SPT (fSPT) states is still not fully resolved for the most general symmetry groups, despite of rapid progresses in recent years~\cite{GuSuH,Kapustin2015,Gaiotto2016,Kapustin2017,Gaiotto2017,QRWangBSuH,JuvenWang2018,Cheng2018,LanFSPT2018X,QRWangFSPT2}.
Physically, the ground state of fSPT phases can be understood in terms of decorated fluctuation domain walls. Mathematically, the fSPT classification is described by generalized cohomology theory which is a combination of layers of cochains in the cohomology of the symmetry group.
These cochains must satisfy certain conditions, which are expressed by obstruction functions mapping cochains to another cocycle~\cite{QRWangFSPT2,Morgan2016,Morgan2018}.
A central task in the classification scheme is to compute such obstruction functions.
For realistic symmetry groups of physical interests, the computational cost of such functions can be high or even prohibitive.
This not only seriously limits the application of the fSPT-classification results to realistic systems, but also inconveniences theoretical studies of fSPT classification by causing difficulties in constructing and studying examples.

In this Letter, we develop an efficient algorithm to make it possible to evaluate obstruction functions for general discrete symmetry groups, based on the physical concept of topological invariants and mathematical techniques in homology algebra. 
In particular, when the outcome of the cocycle function is a cocycle in the bosonic SPT layer (the details of the layers will be reviewed below), the algorithm has an intuitive interpretation as evaluating partition functions on representative space-time manifolds with suitable symmetry-flux insertions~\cite{Tantivasadakarn2017}.
Previously, these partition functions, or equivalent quantities like braiding statistics~\cite{levin12,CJWang2014,CJWang2015,MChengLoopBraid2018,JuvenWang2019X} were used as topological invariants in the process of computing the outcome of cocycle functions.
So far, such invariants have only been constructed for special symmetry groups, e.g., finite Abelian groups.
Our algorithm not only provides a way to efficiently construct and compute these invariants automatically for a large class of symmetry groups, even including infinite ones, but also generalizes them to more tasks, e.g., including the cases where the outcoming cocycle is in other layers of an underlying generalized cohomology theory. For example, our algorithm can be easily generalized for solving similar obstruction functions in the classification of symmetry enriched topological (SET) phases~\cite{XChen2015,Barkeshli2019,BarkeshliRelAnomaly,BarkeshliAbs2020,SQNing2019X}.

Since the techniques of homology algebra we use to develop this algorithm may be unfamiliar to physicists, we include a brief review of related concepts and notations, as well as implementation details
of algorithm stated with standard mathematical notations, in the Supplementary Material (SM), while avoiding the jargons in the main text.
The SM also includes a short introduction to the software package SptSet~\cite{SptSet}, which is written in GAP~\cite{GAP} by one of the authors.
It implements our algorithm and can produce the results listed in Tables~\ref{tab:2dwallpaper} and \ref{tab:2dwallpaper2}.

\paragraph{The fSPT classification.}

We begin with a brief review for the classification of fSPT states we want to compute.
One fruitful scheme for computing the classification is to construct the SPT states using domain-wall decoration~\cite{Gaiotto2017,QRWangFSPT2}.
In this scheme, a ($d+1$)-dimensional SPT state is divided into different layers, where $(p+1)$-dimensional invertible topological orders (iTOs) are decorated onto the $(d-p)$-dimensional symmetry domain walls, respectively.
Here, iTOs are states that do not have any fractionalized excitations, but are still nontrivial even when there is no symmetry to protect it.
For interacting-fermion systems, the invertible topological orders include complex-fermion modes in (0+1)D, Kitaev chains in (1+1)D and $p+ip$-wave topological superconductors in (2+1)D.
Denoting the classification of $(d+1)$-dimensional iTOs by $\iTO^{d+1}$, we have $\iTO^1 = \mathbb Z_2$, $\iTO^2=\mathbb Z_2$, $\iTO^3=\mathbb Z$, respectively~\cite{Ryu2010}.
Using this notation, the decoration of $(p+1)$-dimensional iTOs on $(d-p)$-dimensional symmetry domain walls is classified by a cocycle $n_{d-p}\in H^{d-p}(G_b, \iTO^{p+1})$, where $0\leq p<d$ and $G_b$ denotes the group of bosonic symmetries, which is the quotient group $G_f/\mathbb Z_2^f$, where $G_f$ and $\mathbb Z_2^f$ denotes the total symmetry group of the fermionic system and the fermion-parity symmetry, respectively.
Finally, there is another layer of a bosonic SPT, described by $\nu_{d+1}\in H^{d+1}[G_b, \uone_T]$.

However, the classification of fSPTs is not simply a direct sum of the aforementioned cohomology classes.
In particular, a decoration $n_p$ can be anomalous:
it cannot be realized in a purely $d$-dimensional system, and can only be realized on the boundary of a system in one-higher dimension, with a decoration in a higher layer $n_{p'}$ or $\nu_{d+2}$~\cite{QRWangASPT}.
Mathematically, such a bulk-boundary relation is described by an obstruction function~\cite{Gaiotto2017,QRWangFSPT2}:
\begin{equation}
  \label{eq:Op}
  n_{p'} = O_{p'}[n_p]\text{ or }\nu_{d+2}=O_{d+2}[n_p].
\end{equation}
Physically, this means that the $n_p$ decoration can and must be realized on the surface of the $O_{p'}[n_p]$ decoration.
The application of these obstruction functions is two-fold~\cite{QRWangASPT}:
On one hand, if $O_{d+2}[n_p]$ does not vanish, it signals that $n_p$ does not describe a valid SPT state and should be eliminated.
On the other hand, the corresponding $O_{p'}$ describes a trivial SPT state in one-higher dimension, because its surface can be gapped out without symmetry-breaking or fractionalization.
Hence, the computation of the obstruction functions plays a central role in classifying fSPTs.
After the obstruction-free and nontrivial cocycles are obtained, another subtlety in determining the group structure of fSPT classes is the group-extension problem:
when adding two decorations $n_p$ and $n^{\prime}_p$, if the result is a trivial cocycle in this layer, the physical result may be a nontrivial SPT state in a higher layer.

The obstruction functions and the group-extension functions are both functions mapping one or two $p$-cocycles to a $p'$-cocycle.
We shall use the evaluation of obstruction functions, in particular $O_{d+2}$, as an example in the main part of this work, although the algorithm can be readily applied to other tasks.
In previous works, the obstruction functions are derived in terms of a special form of cocycles, known as the homogeneous or the inhomogeneous cocycles~\cite{QRWangFSPT2}.
In particular, an $n$-cocycle is a function $\alpha(g_1, \ldots, g_n)$, mapping a combination of $n$ group elements (which will be denoted by $[g_1|\cdots|g_n]$), to a complex number in $U(1)$.
Mathematically, it can be viewed as cocycle on a simplicial-complex realization of the classifying space of the symmetry group.
Although this form of cocycles is convenient for theoretical derivation, it is cumbersome for computation.
In particular, the computational complexity of determining the cohomology class of a cocycle computed from an obstruction function scales as $(|G|-1)^{3n}$, where $|G|$ is the order of the group and $n$ is the order of the cocycle.
This complexity quickly becomes prohibitive for complex symmetry groups of physical interests.

\paragraph{The algorithm.}
Using mathematical tools in homology algebra, we construct an algorithm to accelerate this and other similar tasks in the evaluation of the cocycle functions.
Intuitively, this is carried out by mapping the cocycles into a different basis, which can equivalently express all cohomology classes in the group-cohomology theory but is much smaller.

In group-cohomology theory, it is well-known that the cohomology of a group is related to the cohomology of the classifying space of that group,
and therefore can be computed using the chain complex of the classifying space, known as the (free) resolution associated with the group~\cite{Joyner2007,brown2012cohomology}.
The resolution provides the basis for writing down the cocycles.
For a given group, there are different choices of realizations of the classifying space, resulting in different resolutions, and all choices are homotopically equivalent to each other, meaning that the resulting cohomology group is independent of the choice.
In particular, the well-known inhomogeneous cocycles correspond to a particular simplicial construction of the classifying space, with the resolution known as the bar resolution in mathematics.
However, not all resolutions are created equally: some resolutions are smaller than others, meaning that there are fewer basis for expressing the cocycles in each dimension, and therefore require much less computational resources in practice.
The bar resolution, on the other hand, is one of the biggest resolutions.
Unfortunately, some of the obstruction functions are only known, to the best of our knowledge, in terms of the inhomogeneous cocycles or the bar resolution, which is indeed convenient for theoretically deriving the obstruction functions due to the simplicial structure of the corresponding classifying space.
Hence, we propose an algorithm for accelerating the evaluation of the cocycle functions, by first converting the input cocycle to an inhomogeneous cycle, computing the resulting cocycle using the inhomogeneous-cocycle formula, then converting the result back to a cocycle in a smaller resolution.
Since the most time-consuming step is to find the cohomology class of a cocycle,
this step becomes much faster in the smaller resolution, providing an overall acceleration to the whole process.

In particular, in a resolution with only a few basis elements, the task is simplified to evaluating a few topological invariants using entries in the inhomogeneous cocycle.
Physically, when the coefficient of the cohomology is U(1), these invariants can be viewed as partition functions on space-time manifolds with nontrivial symmetry fluxes.
Previously, such invariants have been constructed case-by-case for simple cases~\cite{CJWang2015,MChengLoopBraid2018,Tantivasadakarn2017}.
Our algorithm generalizes and automates the construction of these invariants to a large class of groups.
We notice that similar techniques have been used in mathematics in computing properties of higher groups~\cite{Ellis2012}.

To be more concrete, for a discrete group $G$, we construct two sets of basis (known as resolutions) representing the same set of cohomology classes: the bar resolution whose basis elements are $[g_1|\cdots|g_n]$, and a simplified solution whose basis elements are denoted abstractly by $e^n_1,\ldots,e^n_{r_n}$.
The core task of our algorithm is to construct two maps between the two bases: $f$ maps each element in the basis of the simplified resolution to the bar resolution,
\begin{equation}
  \label{eq:f-def-intro}
  f(e^n_i) = \sum_{g_1,\ldots,g_n}\phi_i(g_1,\ldots,g_n)[g_1|\cdots|g_n];
\end{equation}
$g$ maps each basis in the bar resolution to the simplified resolution,
\begin{equation}
  \label{eq:g-def-intro}
  g([g_1|\cdots|g_n]) = \sum_i\gamma_i(g_1,\ldots,g_n)e^n_i.
\end{equation}
Here, for certain $i$ and $[g_1|\cdots|g_n]$, the coefficients $\phi$ and $\gamma$ actually belong to the integral group ring $\mathbb ZG$ reviewed in Sec.~VI of the SM.

Using the maps $f$ and $g$, we can compute the obstruction function $O_{d+2}$ for cocycles in the simplified basis as the following.
Given a $p$-cocycle in the simplified basis, represented as $\alpha(e^p_i)$, we first convert it to an inhomogeneous cocycle, denoted as $\bar\alpha$,  using the map $g$:
\[\bar\alpha(g_1,\ldots,g_n)=\alpha(g([g_1|\cdots|g_n])).\]
The obstruction function $O_{d+2}$ is then computed using the formula for inhomogeneous cocycles.
The result of this formula is an inhomogeneous $(d+2)$-cocycle, denoted by $\bar\beta = O_{d+2}[\bar\alpha]$.
$\bar\beta$ is then converted to the simplified basis using the map $f$, as
$\beta(e^{d+2}_i)=\bar\beta(f(e^{d+2}_i))$.
We notice that, to convert cocycles in the simplified basis (bar-resolution basis) to ones in the bar-resolution basis (simplified basis), we use the map $g$ ($f$), respectively.
This is because the cocycles, analogous to linear functions in linear spaces, are contravariant under the changes of basis.
Mathematically, the conversion between two types of cocycles are known as pullback maps induced by the chain maps $g$ and $f$.

\paragraph{Construct the chain maps.}
The maps $f$ and $g$ are known as chain maps between the two resolutions, and can be constructed using standard homology-algebra techniques, which are outlined here.
Details of the algorithm and reviews of standard notation in homology can be found in Sec. II of the SM.
Without losing generality, we consider constructing a chain map $f$ between two resolutions $F$ and $F'$, whose bases in dimension $n$ are denoted by $e_i^n$ and $e_i^{n\prime}$, respectively.
The construction is iterative:
We start with the lowest dimension $n=0$, where both resolutions contain only one basis and the construction of the map is obvious.
Then, assuming the map is constructed for dimension $n-1$, we now proceed to dimension $n$ and compute $f(e_i^n)$ for each basis $e_i^n$.
In order to preserve the algebraic structure, it is required that the chain map $f$ commutes with the boundary operators $\partial$ and $\partial'$ of the two resolutions $F$ and $F'$.
(The boundary can be viewed as the dual operation of the cobounary operator on the cochains.)
Hence, we require that
\[\partial' f(e_i^n) = f(\partial e_i^n).\]
Notice that the right hand side of the equation can be computed using the map constructed for dimension $n-1$.
A proper choice of $f(e_i^n)$ can then be computed using a contracting homotopy $s'$ of $F'$:
\[f(e_i^n) = s'[f(\partial e_i^n)].\]
Intuitively, a contracting homotopy can be viewed as an ``inverse'' of the boundary operator $\partial'$, and it is essential to our algorithm.
For the inhomogeneous cocycles or the bar resolution, a standard choice of contracting homotopy is reviewed in the SM.
For the simplified resolutions, a contracting homotopy must be constructed along with the resolution.

\paragraph{Construct the resolution.}
For a large class of groups, including all finite groups and the space groups, a simplified resolution suitable for our algorithm, accompanied by a contracting homotopy, can be constructed using the procedures introduced in Ref.~\cite{Ellis2004}, which is implemented by the HAP package~\cite{HAP} in the GAP software~\cite{GAP}.
This includes all 17 2D wallpaper groups and most 3D space groups.

Moreover, when a group $G$ is expressed as an extension of $Q$ by $N$, a resolution of $G$ can be constructed using resolutions of $Q$ and $N$~\cite{Wall1961}.
Compared to the construction in Ref.~\cite{Ellis2004}, this method is easier to implement.
This method can also be applied to all 2D wallpaper groups.
It is well-known that a 2D wallpaper group can be viewed as an extension of a point group $P$ by the translation-symmetry group $T=\mathbb Z^2$: $P=G/T$.
Hence, we can construct a resolution over $G$ from resolutions over $P$ and $\mathbb Z^2$, using Wall's construction~\cite{Wall1961}.
$\mathbb Z^2$ has a simple resolution because $B\mathbb Z^2$ is simply the 2D torus $T^2$.
For the point group $P$, we recall that there are eight possible nontrivial point groups in 2D: $C_{2,3,4,6}$ and $D_{2,3,4,6}$.
For the cyclic groups, simple resolutions over $C_n=\mathbb Z_n$ are reviewed in Sec.~II of the SM.
For the dihedral groups, they can in turn be expressed as split extensions $D_n=C_n\rtimes\mathbb Z_2$.
Hence, a simple free resolution can be constructed again using Ref.~\cite{Wall1961}.
Therefore, finite-rank free resolutions over 2D wallpaper groups can be constructed by combining simple resolutions over $\mathbb Z_n$ and $\mathbb Z$ using Ref.~\cite{Wall1961}.
We note that this approach also works for 3D space groups.

\paragraph{Example of the $\mathbb Z_{2n}$ symmetry group.}
Here, we sketch a simple example for SPT phases protected by the $G=\mathbb Z_{2n}$ group.
For concreteness, we demonstrate how to simplify the computation of $O_4$ obstruction for a 3D SPT state with a Majorana-chain decoration, denoted by $n_2$.
This obstruction represents the failure of finding a complex-fermion decoration, and its formula is given by Eq.~(38) in  Ref.~\cite{QRWangBSuH}, for the simple case where the unitary $\mathbb Z_{2n}$ symmetry extends trivially over the fermion-parity symmetry.

As explained in the SM, a simplified resolution for $G=\mathbb Z_{2n}$ consists of only a single basis in each dimension $k$, denoted by $e_k$.
Using this resolution, the single nontrivial choice of $n_2$ is given by $n_2(e_2)=1$.
The obstruction cocycle, $\alpha=O_4[n_2]$, is also represented by a single entry $\alpha(e_4)$.
To compute this, we first express $\alpha$ by an inhomogeneous cocycle using the mapping $f$ in Eq.~\eqref{eq:f-def-intro}, which maps $e^4$ to
$f(e^4)=\sum_{i,j=1}^{2n-1}[a^i|a|a^j|a].$
[See Eq.~(21) in Sec.~IIB of SM.]
This expresses $\alpha(e^4)$ as a linear combination of inhomogeneous cocycles
$\alpha(e^4)=\sum_{i,j=1}^{2n-1}\bar\alpha(a^i, a, a^j, a)$.
Using the equation for $O_4$ obstruction in inhomogeneous cocycles, $\bar\alpha =O_4[n_2]=\bar n_2\cup \bar n_2$, we obtain
\begin{equation}
  \label{eq:example:alpha}
  \alpha(e^4)=\sum_{i,j=1}^{2n-1}\bar n_2(a^i, a)\bar n_2(a^j, a).
\end{equation}
Next, we compute the entries of the inhomogeneous cocycle $\bar n_2$ from $n_2$, using the chain map $g$ in Eq.~\eqref{eq:g-def-intro}.
Using the form of $g$ in Eq.~(24) of the SM, we can reach $\bar n_2(a^i, a)=\delta_{i, 2n-1}$.
Plug this into Eq.~\eqref{eq:example:alpha}, we obtain $\alpha(e^4)=1$, indicating that the $O_4$ obstruction is nontrivial.
Therefore, such a Majorana-chain decoration is obstructed.

Using the simplified resolution, determining $\alpha=O_4$ only needs to compute the $(2n-1)^2$ entries in Eq.~\eqref{eq:example:alpha}.
Therefore, the computational cost increases with $n$ as $O[(2n-1)^2]$.
On the other hand, in traditional methods, not only do we need to compute all $(2n-1)^4$ entries in $\bar\alpha$, but we also need to solve a set of linear equations (with integral coefficients) with dimension $(2n-1)^4$-by-$(2n-1)^4$, and the associated computational cost is $O[(2n-1)^{12}]$.
Therefore, the algorithm presented here greatly reduces the computational cost.

\paragraph{Solving a twisted cocycle equation.}
This is another computationally heavy task that needs to be and can be accelerated using a simplified resolution.
In the computation of fSPT classification, when an obstruction function in Eq.~\eqref{eq:Op} is a trivial but nonvanishing cocycle, the cochain representing the decoration $n_{p'-1}$ or $\nu_{d+1}$ must satisfy the following twisted cocycle equation~\cite{QRWangBSuH,QRWangFSPT2},
\begin{equation}
  \label{eq:dn}
  dn_{p'-1} = O_{p'}[n_p] \text{ or } d\nu_{d+1} = O_{d+2}[n_p].
\end{equation}
The fact that the r.h.s. of this equation is trivial guarantees that the equation has solutions, and the task is to seek one particular solution.
Using the inhomogeneous cochains or the bar resolution, this task is also time-consuming as the matrix form of Eq.~\eqref{eq:dn} has dimension $(|G|-1)^{p'-1}\times(|G|-1)^{p'}$ or $(|G|-1)^{d+1}\times(|G|-1)^{d+2}$.
Here, we briefly sketch the idea of accelerating this using a simplified resolution, and the details of the implementation can be found in Sec.~III of the SM.

To simplify the notations, we consider the generic problem of finding one solution of the cocycle equation in terms of inhomogeneous cochains,
\begin{equation}
\label{eq:db=a}
d\bar\beta = \bar\alpha,
\end{equation}
Naively, one may try to solve Eq.~\eqref{eq:db=a} by mapping $\bar\alpha$ to a cochain in the simplified resolution using the map $f$ as $\alpha$, find a solution $\beta'=d\alpha$ there, and map it back to an inhomogeneous cocycle using the map $g$, which we denote by $\bar\beta'$.
However, the inhomogeneous cocycle $\bar\beta'$ constructed this way is not a solution of Eq.~\eqref{eq:db=a}, because the two chain maps $f$ and $g$ are not the inverse of each other.
Instead, $fg$ is only homotopic to the identity map, meaning that it can be related to identity using a homotopy $h: fg\sim\id$.
In fact, the homotopy $h$ can be used to construct a ``correction'' cochain we denote by $h^\ast(\bar\alpha)$:
$h^\ast(\bar\alpha)(g_1,\ldots,g_n) = \bar\alpha(h([g_1|\cdots|g_n]))$, and one solution of Eq.~\eqref{eq:db=a} is then given by $\bar\beta = \bar\beta' - h^\ast(\bar\alpha)$.
As described in details in the SM, the homotopy $h$ can be viewed as an analogue of the chain maps $f$ and $g$ (actually, it is a degree-1 map from the bar resolution to itself), and can be computed using a similar iterative procedure.

\paragraph{Wallpaper-group SPT classification}

To demonstrate the power of our algorithm, we compute the classification of 2D fSPTs protected by the 17 2D wallpaper groups~\cite{wg}.
This task is impossible using the inhomogeneous cocycles, because the wallpaper groups are infinite.
However, using our algorithm, it becomes a finite problem and can be solved using a computer program.

The problem we consider is the 2D fSPTs protected by an onsite symmetry group $G$, which has the same group structure as one of the 17 wallpaper groups.
We assume that the proper and improper operations in $G$ act as unitary and antiunitary operations, respectively.
Furthermore, we assume that the total symmetry group is a direct product of the wallpaper group $G$ and the fermion-parity symmetry group $\mathbb Z_2^f$: $G_f=G\times\mathbb Z_2^f$.
According to the crystalline equivalence principle~\cite{ThorngrenElse2018, MCheng2018X}, this fSPT classification is the same as the classification of topological crystalline states protected by the wallpaper group $G$, formed by fermions transforming projectively under $G$ as the physical spin-$\frac12$ electrons do.
Therefore, the classification we compute here can also guide the search of such topological crystalline states on 2D lattices.
The results we obtain are listed in Table~\ref{tab:2dwallpaper}.
Moreover, we also compute the classification for spin-$\frac12$ fermions, which corresponds to topological crystalline states formed by spinless fermions, and the results are listed in Table~\ref{tab:2dwallpaper2}.
We note that they agree with recent results obtained by real-space constructions of the corresponding topological crystalline states~\cite{JHZhangUnpub}.

\begin{table}
  \caption{Classification of 2D fSPTs protected by 2D wallpaper groups, where fermions are spinless (spin-1/2 if the symmetry group is treated as spatial symmetries.)
  The answer is listed in terms of the Majorana-chain (MC), complex-fermion (CF), and bosonic (B) layers. The total number of phases combining all three layers is also listed in the last column. The overall group structure is not computed in this paper.}
  \label{tab:2dwallpaper}
  \begin{tabularx}{\columnwidth}{CCCCC}
    \hline\hline
    SG & MC & CF & B & Total\\
    \hline
    p1 & $2\mathbb Z_2$ & $\mathbb Z_2$ & 0 & 8 \\
    p2 & $3\mathbb Z_2$ & $4\mathbb Z_2$ & $4\mathbb Z_2$ & 2048\\
    p1m1 & $\mathbb Z_2$ & $2\mathbb Z_2$ & $2\mathbb Z_2$ & 32\\
    p1g1 & $2\mathbb Z_2$ & $\mathbb Z_2$ & 0 & 8\\
    c1m1 & $\mathbb Z_2$ & $\mathbb Z_2$ & $\mathbb Z_2$ & 8\\
    p2mm& 0 & 0 & $8\mathbb Z_2$ & 256\\
    p2mg& $2\mathbb Z_2$ & $3\mathbb Z_2$ & $3\mathbb Z_2$ & 256\\
    p2gg& $2\mathbb Z_2$ & $2\mathbb Z_2$ & $2\mathbb Z_2$ & 64\\
    c2mm& $\mathbb Z_2$ & $\mathbb Z_2$ & $5\mathbb Z_2$ & 128\\
    p4& $2\mathbb Z_2$ & $3\mathbb Z_2$ & $\mathbb Z_2\oplus2\mathbb Z_4$ & 1024\\
    p4mm& 0 & 0 & $6\mathbb Z_2$ & 64\\
    p4gm& $\mathbb Z_2$ & $\mathbb Z_2$ & $2\mathbb Z_2\oplus\mathbb Z_4$ & 64\\
    p3& 0 & $\mathbb Z_2$ & $3\mathbb Z_3$ & 54\\
    p3m1& 0 & $\mathbb Z_2$ & $\mathbb Z_2$ & 4\\
    p31m& 0 & $\mathbb Z_2$ & $\mathbb Z_6$ & 12\\
    p6& $\mathbb Z_2$ & $2\mathbb Z_2$ & $2\mathbb Z_6$ & 288\\
    p6mm& 0 & 0 & $4\mathbb Z_2$ & 16\\
    \hline\hline
  \end{tabularx}
\end{table}

\begin{table}
  \caption{Classification of 2D fSPTs protected by 2D wallpaper groups, where fermions carry spin-1/2 (spinless if the symmetry group is treated as spatial symmetries.) The answer is listed in terms of the Majorana-chain (MC), complex-fermion (CF), and bosonic (B) layers. The total number of phases combining all three layers is also listed in the last column. The overall group structure is not computed in this paper.}
  \label{tab:2dwallpaper2}
  \begin{tabularx}{\columnwidth}{CCCCC}
    \hline\hline
    SG & MC & CF & B & Total\\
    \hline
    p1 & $2\mathbb Z_2$ & $\mathbb Z_2$ & 0 & 8 \\
    p2 & 0 & $3\mathbb Z_2$ & $\mathbb Z_2$ & 16\\
    p1m1 & $2\mathbb Z_2$ & $3\mathbb Z_2$ & $\mathbb Z_2$ & 64\\
    p1g1 & $2\mathbb Z_2$ & $\mathbb Z_2$ & 0 & 8\\
    c1m1 & $2\mathbb Z_2$ & $\mathbb Z_2$ & $\mathbb Z_2$ & 16\\
    p2mm& 0 & $4\mathbb Z_2$ & $4\mathbb Z_2$ & 256\\
    p2mg& $\mathbb Z_2$ & $3\mathbb Z_2$ & $\mathbb Z_2$ & 32\\
    p2gg& $\mathbb Z_2$ & $\mathbb Z_2$ & $\mathbb Z_2$ & 8\\
    c2mm& 0 & $3\mathbb Z_2$ & $2\mathbb Z_2$ & 32\\
    p4& 0 & $2\mathbb Z_2$ & $\mathbb Z_2\oplus\mathbb Z_4$ & 32\\
    p4mm& 0 & $4\mathbb Z_2$ & $3\mathbb Z_2$ & 128\\
    p4gm& 0 & $2\mathbb Z_2$ & $2\mathbb Z_2$ & 16\\
    p3& 0 & $\mathbb Z_2$ & $3\mathbb Z_3$ & 54\\
    p3m1& $\mathbb Z_2$ & $\mathbb Z_2$ & $\mathbb Z_2$ & 8\\
    p31m& $\mathbb Z_2$ & $\mathbb Z_2$ & $\mathbb Z_6$ & 24\\
    p6& 0 & $\mathbb Z_2$ & $\mathbb Z_3\oplus\mathbb Z_6$ & 36\\
    p6mm& 0 & $2\mathbb Z_2$ & $2\mathbb Z_2$ & 16\\
    \hline\hline
  \end{tabularx}
\end{table}

\paragraph{Conclusion.}
\label{sec:conclusion}
In this Letter, we have given an algorithm to accelerate the computation of certain maps between different cohomology classes of a symmetry group, which is a common and essential task in classifying fSPTs.
Using the fact that the same cohomology classes can be obtained from different choices of classifying space of the group, or algebraically different resolutions,
the algorithm constructs chain-maps between the standard choice of resolution, where the formula of the desiring maps are known, and a simplified choice of resolution where the computation is much easier.
Such chain-maps then allow us to convert cocycles between two choices of resolutions, and to simplify the computation of the maps between cohomology classes.

Our algorithm not only reproduces some known results on finite groups with a faster speed, but also works for infinite discrete groups, like the 2D wallpaper groups and 3D space groups.
Hence, it can be used to compute examples for the study of fSPT classification, and to compute fSPT classification for symmetry groups relevant to materials, which usually include space-group symmetries.
Furthermore, recent progresses on the classification of 2D symmetry-enriched topological (SET) states~\cite{XChen2015,Barkeshli2019,BarkeshliRelAnomaly,BarkeshliAbs2020} and 3D U(1) quantum spin liquids~\cite{SQNing2019X} also involve obstruction functions that map between cohomology classes of the symmetry group.
The computation of these obstruction functions can also be accelerated by our algorithm.

\begin{acknowledgements}
YQ is grateful to Hao Zheng, Chenjie Wang and Yang Su for invaluable discussions.
YQ acknowledges support from National Science Foundation of China under grant number 11874115. ZCG acknowledges support from Hong Kong's Research Grants Council under grant GRF No.14306918 and ANR/RGC Joint Research Scheme No. A-CUHK402/18.
\end{acknowledgements}

\bibliography{cohomology,spt}

\end{document}


\title{Supplementary Materials}
\author{Yunqing Ouyang}
\affiliation{State Key Laboratory of Surface Physics, Fudan University, Shanghai 200433, China}
\affiliation{Center for Field Theory and Particle Physics, Department of Physics, Fudan University, Shanghai 200433, China}
\author{Qing-Rui Wang}
\thanks{Present address: Yau Mathematical Sciences Center, Tsinghua University, Beijing 100084, China}
\affiliation{Department of Physics, Yale University, New Haven, CT 06511, USA}
\author{Zheng-Cheng Gu}
\email{zcgu@phy.cuhk.edu.hk}
\affiliation{Department of Physics, The Chinese University of Hong Kong, Shatin, New Territories, Hong Kong, China}
\author{Yang Qi}
\email{qiyang@fudan.edu.cn}
\affiliation{State Key Laboratory of Surface Physics, Fudan University, Shanghai 200433, China}
\affiliation{Center for Field Theory and Particle Physics, Department of Physics, Fudan University, Shanghai 200433, China}
\affiliation{Collaborative Innovation Center of Advanced Microstructures, Nanjing 210093, China}
\date{\today}

\maketitle

\tableofcontents

\section{Motivation: $\mathbb Z_n$-SPT in 2D}
\label{sec:motivation}

We first motivate our method using the task of identifying the cohomology class of a cocycle, with the example of a simple cyclic group $\mathbb Z_n$.
In this case, cocycles in the simplified basis are related to the topological invariants used in previous studiess~\cite{Tantivasadakarn2017}.

In particular, we consider the task of checking whether the result of an obstruction function is a trivial or nontrivial cocycle.
(For the definition of cocycles and their cohomology classes, see Sec.~\ref{sec:bg}.)
Instead of using coboundary equations for the inhomogeneous cochains directly, which results in a large computational cost, we construct topological invariants.
In general, a $(d+1)$-cocycle $\alpha\in H^{d+1}[G,\uone]$ can be interpretted as a $d$-dimensional bSPT state.
In fact, such a cocycle can be used to construct partition functions on any closed $(d+1)$D manifold $M$, with arbitrary symmetry fluxes of $G$ inserted in noncontractible loops of $M$~\cite{DijkgraafWitten}.
Such a combination of closed (3+1)D manifold and symmetry fluxes is knwon as a $G$-bundle.
A trivial 4-cocycle, representing a trivial SPT phase, gives a partition function that evaluates to the trivial value of $+1$ on any $G$-bundle; a nontrivial cocycle, on the other hand, evaluates to nontrivial values on some nontrivial $G$-bundles.

Furthermore, only a few number of representative $G$-bundles need to be checked, each detecting one root cohomology class in $H^{d+1}[G,\uone]$.
If the partition function is trivial on all these $G$-bundles, the corresponding cocycle is trivial.

\begin{figure}[htbp]
\centering
  \subfigure[\label{fig:lens:cw}]{\includegraphics{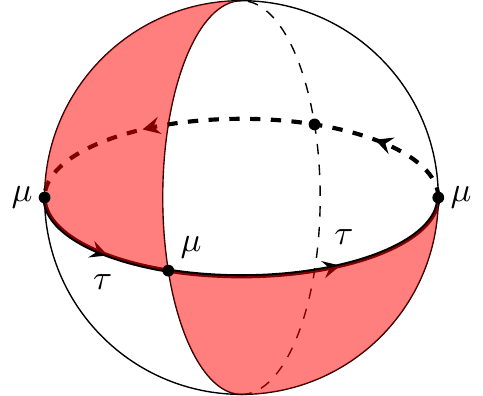}}
  \subfigure[\label{fig:lens:d}]{\includegraphics{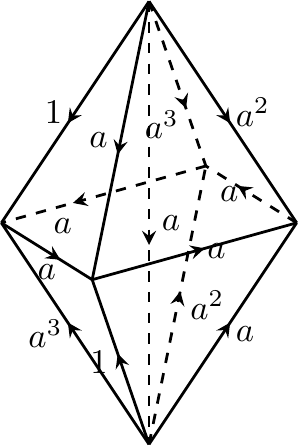}}
  \caption{\label{fig:lens}The lens space $L_4(1)$.
  (a) $L_4(1)$ constrcuted by gluing the upper and lower hemispheres of a 3-ball. The upper hemisphere is first rotated by 90 degrees before glued to the lower hemisphere, as indicated by the red marks on both hemispheres, which are glued together.
  $L_4(1)$ has one 3-cell, which is the interior of the ball; it has one 2-cell, which is the upper hemisphere or the lower hemisphere (they are identified during the gluing); it has one 1-cell, which is one of the segments $\tau$ on the equator shown in thick lines with arrows (the equator is divided into four segments, which are identified with each other after the gluing); it has one 0-cell, which is the starting and the ending point of $\tau$ and labeled by $\mu$.
  (b) A triangulation ($\Delta$-complex decomposition) of $L(4, 1)$, and a flat connection realzing the nontrivial symmetry flux along $\tau$.}
\end{figure}

To demonstrate this procedure, we consider (2+1)D manifolds, which are easy to illustrate, and a simple symmetry group: the cyclic group $G=\mathbb Z_n=\langle a|a^n=1\rangle$.
This example also appeared in Ref.~\cite{Tantivasadakarn2017}.
For this simple case, the (2+1)D bSPT phases are classified by $H^3[G,\uone] = Z_n$.
Hence, there is only one root state, and one corresponding representative $G$-bundle.
This $G$-bundle is illustrated in Fig.~\ref{fig:lens} for the case of $n=4$.
The base manifold of this $G$-bundle is constructed by starting from
a solid 3-ball and gluing the two hemispheres on the surface of the ball in the following twisted way:
the upper hemisphere is rotated by an angle of $2\pi/n$, reflected with respect to the equator, and glued to the lower hemisphere.
Consistent with this gluing, the equator can be divided evenly into $n$ segments, which are identified with each other.
Consequently, the starting and end points of each segment are also identified as the same point, and the segment becomes a noncontractible loop.
The gluing creates a closed 3-manifold $M$, which is known as the lens space $L_n(1)$ in mathematics~\cite{HatcherAT}.
This manifold has a nontrivial first homotopy group $\pi_1(M)=\mathbb Z_n$, generated by the noncontractible loop $\tau$ shown on Fig.~\ref{fig:lens:cw}.
The $G$-bundle has a nontrivial symmetry flux $a$ ($a$ labels the generator of the $\mathbb Z_n$ group) along this loop.

We now evaluate on this $G$-bundle the partition function constructed from a 3-cocycle $\alpha\in H^3[G,\uone]$.
We assume that $\alpha$ is computed as an inhomogeneous cocycle.
As explained in details in 
Sec.~\ref{app:inhomogeneous},
an inhomogeneous 3-cocycle can be used to construct partition functions on a simplicial complex with a flat gauge connection, which is basically a triangulated space consists of many tetrahedra (3-simplices).
The gauge connection consists of $g_{ij}\in G$ assigned to each edge $[v_iv_j]$ in the complex, satisfying two constraints:
First, the total flux going around a triangle $[v_iv_jv_k]$ must vanish: $g_{ij}g_{jk}=g_{ik}$.
Second, the total flux going around a noncontractible loop in $\pi_1(M)$ must produce the assigned symmetry flux in the $G$-bundle.
On such a simplicial-complex realization of the $G$-bundle, a partition function of the SPT phase represented by the cohomology class $\alpha$ is constructed by multiplying weights associated with each tetrahedron:
on one tetrahedron, denoted by its four vertices as $[v_0v_1v_2v_3]$, the weight is given by
\begin{equation}
\exp\left\{\pm2\pi i\langle\alpha,[v_0v_1v_2v_3]\rangle\right\}
=\exp\left\{\pm2\pi i\alpha(g_{01},g_{12},g_{23})\right\}.
\end{equation}
Here, the overall sign in the phase is plus (minus) if the orientation of the simplex is positive (negative), respectively.

Therefore, in order to evaluate the partition function on the $G$-bundle in Fig.~\ref{fig:lens:cw}, we must first decompose it into a simplicial complex, and assign a choice of flat connection $g_{ij}$.
One particular construction is given in Fig.~\ref{fig:lens:d}.
It is then straightforward to compute the partition function:
\begin{equation}
\label{eq:Zlens}
Z = \exp\left\{2\pi i
\sum_{j=1}^n\alpha(a, a^j,a)\right\}.
\end{equation}

This partition function detects the classification of the SPT states:
the trivial SPT phase gives $Z=+1$, while the root state of nontrivial SPTs gives $Z=e^{i2\pi/n}$.
Therefore, it can be used as a topological invariant to determine the cohomology class of the cocycle $\alpha$.
In general, the value of $Z$ can be $Z=e^{i2\pi k/n}$, where $k=0, 1,\ldots,n-1$ indicates the cohomology class of $\alpha$.
In practice, the cohomology class of an inhomogeneous cocycle can be determined by evaluating such topological invariants instead of solving the cocycle equations of the inhomogeneous cocycles, which is a time-consuming task.
In the rest part of the paper, we will introduce automated procedures to construct such topological invariants for generic discrete groups.

\section{Algebraic description}
\label{sec:algebraic}


In this section, we describe a general algorithm for constructing topological invariants.
Physically, the topological invariants are constructed from evaluating the partition functions on representative $G$-bundles.
Hence, the algorithm contains two parts: First, one chooses representative $G$-bundles by constructing the classifying space of $G$, denoted by $BG$.
Second, a triangulation of the $G$-bundles is computed by constructing a cellular map from $BG$ to a standard simplicial realization of $BG$.
These two steps are discussed in Secs.~\ref{sec:bg} and ~\ref{sec:chainmap}, respectively.
Finally, in Sec.~\ref{sec:coboundary}, we combine the two steps and construct an algorithm for constructing the invariants that check the trivialness of a cocycle.
Although the algorithm has a nice interpretation in terms of evaluating SPT partition functions on $G$-bundles,
the derivation of the chain map can be described purely algebraically.
For conciseness, we only discuss the algebraic construction of the algorithm in this section, and defer the discussion of physical interpretation to Sec.~\ref{sec:physical}.

\subsection{Classifying space and resolution}
\label{sec:bg}

The classifying space of $G$, denoted by $BG$, is a topological space satisfying the following conditions: its first homotopy group (or the fundamental group) is $\pi_1(BG) = G$, and all its higher homotopy groups vanish: $\pi_k(BG)=0$, $k>1$.
Closely related to $BG$, the universal bundle $EG$, is also the universal cover of $BG$.
Since the fundamental group of $BG$ is $G$, $EG$ can be viewed as a topological space with a free action of $G$, and $BG$ is the quotient space $BG = EG/G$.

$EG$ is called the universal bundle, because any $G$-bundle can be constructed as a pullback bundle from its base space $BG$.
As a result, a cohomology class on $BG$ can be used to define partition functions on all possible $G$-bundles.
This leads to the conclusion that $d$-dimensional bSPT phases are classified by $H^{d+1}[BG, \uone]=H^{d+1}[G,\uone]$.
In fact, in the real computation, we do not need all geometric details of $BG$ and $EG$.
Instead, only the cellular chain complex of $EG$ is needed.
Here, we review the algebraic structure of this chain complex, which is also known as a free $\mathbb ZG$-resolution (of $\mathbb Z$).

Mathematically, we construct $EG$ as a CW-complex, which is a model of topological spaces widely used in algebraic topology, especially in  the theory of singular homology.
The precise definition of a CW-complex can be found in Appendix A of Ref.~\cite{HatcherAT}.
Roughly speaking, a CW-complex is made by gluing cells of different dimensions, where each $d$-dimensional cell, or a $d$-cell for short, is homeomorphic to a $d$-dimensional disk.
We denote the collection of $d$-cells in the CW-complex $EG$ as $(EG)_d$.

In singular-homology theory, a $d$-chain is a formal summation of $d$-cells, with integral coefficients.
Hence, the space of $d$-chains, denoted by $C_d(EG)$, is a $\mathbb Z$-module with basis in $(EG)_d$.
Since $G$ has a free action on $EG$, the modules $C_d(EG)$ are actually free $\mathbb ZG$-modules.
Furthermore, they form the following long exact sequence under the boundary map,
\begin{widetext}
\begin{equation}
\label{eq:chaincomplex}
\cdots\rightarrow
C_k(EG)\xrightarrow{\partial_k}
C_{k-1}(EG)\rightarrow\cdots\rightarrow
C_1(EG)\xrightarrow{\partial_1}
C_0(EG)\xrightarrow{\epsilon}\mathbb Z\rightarrow0.
\end{equation}
\end{widetext}
This long exact sequence is known as the augmented chain complex of $EG$.

In practice, we only need to keep track of the algebraic structure of the chain complex above.
From this view point, we have free $\mathbb ZG$-modules $F_d=C_d(EG)$ forming a long exact sequence,
\begin{equation}
\label{eq:res}
\cdots\rightarrow
F_k\xrightarrow{\partial_k}
F_{k-1}\rightarrow\cdots\rightarrow
F_1\xrightarrow{\partial_1}
F_0\xrightarrow{\epsilon}\mathbb Z\rightarrow0.
\end{equation}
This is called an augmented free $\mathbb ZG$-resolution.

The exactness of the sequences in Eqs.~\eqref{eq:chaincomplex} and \eqref{eq:res} follows the fact that the space $EG$ is contractible.
Mathematically, this means that the identity map
from $EG$ to itself is homotopic to the zero map that maps $EG$ to an empty space.
Such a homotopy equivalence between these two maps is called a contracting homotopy, and it plays an essential role in the construction of chain maps in Sec.~\ref{sec:chainmap}.
Algebraically, a contracting homotopy $s$ is a collection of $\mathbb Z$-linear maps from each module $F_k$ to the module in one higher dimension, $F_{k+1}$, as shown in the following diagram:
\begin{equation}
\begin{tikzcd}
F_{k+1} \arrow[r,"\partial_{k+1}"]
&F_k\arrow[r,"\partial_k"]
\arrow[d,"\id"]
\arrow[ld,"s_k"]&
F_{k-1}\arrow[ld,"s_{k-1}"]\\
F_{k+1} \arrow[r,"\partial_{k+1}"]& F_k\arrow[r,"\partial_k"]&
F_{k-1}
\end{tikzcd}
\end{equation}
This is not a commutative diagram.
Instead, the maps satisfy the following condition,
\begin{equation}
  \label{eq:pssp}
\partial_{k+1}s_k + s_{k-1}\partial_k = \id.
\end{equation}
In other words, the anticommutator between $s$ and $\partial$ is $\id$, which can be understood as the difference between the identity map and the zero map.
Hence, this indicates that $s$ is a homotopy between these two maps.
We emphasize that $s$ is $\mathbb Z$-linear but not $\mathbb ZG$-linear in general, meaning that it does not commute with group action: $s_k(g\cdot x) \neq g\cdot s_k(x)$.

Algebraically, Eq.~\eqref{eq:pssp} implies that $s$ can be viewed as an ``inverse'' of the boundary map:
For a closed $k$-chain $x\in F_k$, the condition $\partial_k x = 0$ simplifies Eq.~\eqref{eq:pssp} to $\partial_{k+1} s_k(x) = x$.
Hence, $s_k(x)$ is a $(k+1)$-chain that borders $x$.
This immediately proves the exactness of the sequence in \eqref{eq:res}, because every cycle $x$ is a boundary of $s_k(x)$.
This operation of finding the inverse of the boundary map using a contracting homotopy will also play a vital role in the construction of chain maps in Sec.~\ref{sec:chainmap}.

Once a resolution is constructed for a group $G$, it can be used to compute the group-cohomology classification and the invariants of the cocycles.
The $k$-cochains are defined as $\mathbb ZG$-linear maps from $F_k$ to the coefficient module $M$, and space of $k$-cochains is denoted by $C^k(G, M)=\hhom_G(F_k, M)$.
Here, the subscript $G$ indicates that the cochains are invariant under the action of $G$:
\begin{equation}
  \label{eq:homg}
  \langle\alpha, gx\rangle = g\langle\alpha,x\rangle.
\end{equation}
In this paper, we use greek letters to denote cochains.
The bracket $\langle\alpha,x\rangle$ denotes evaluating the linear map $\alpha$ on the element $x\in F_k$.
The result of the bracket is a coefficient $\langle\alpha,x\rangle\in M$, and $g \langle\alpha,x\rangle$ denotes the $G$-action on $M$.

The boundary map $\partial_k:F_k\rightarrow F_{k-1}$ naturally induces a coboundary map $d^{k-1}:C^{k-1}(G, M)\rightarrow C^k(G, M)$:
\begin{equation}
\label{eq:cobdry}
\langle d^{k-1}\alpha,x\rangle
=\langle\alpha,\partial_kx\rangle.
\end{equation}
Using the coboundary maps, we can define the $k$-cocycles, which are $k$-cochains satisfying $d^k\alpha = 0$, and the $k$-coboundaries, which are the coboundary of $(k-1)$-cochains, $\alpha=d^{k-1}\beta$.
The spaces of $k$-cocycles and $k$-coboundaries are $Z^k(G, M)=\ker d^k$ and $B^k(G, M)=\img d^{k-1}$, respectively.
The property that $\partial_k\partial_{k+1}=0$, or the boundary of a boundary is empty, implies that $d^kd^{k-1}=0$.
This ensures that $B^k(G, M)$ is a submodule of $Z^k(G, M)$, and allows us to define the $k$-th cohomology of $G$ as the quotient of the two modules,
\begin{equation}
\label{eq:def-Hk}
H^k(G, M)=\frac{Z^k(G, M)}{B^k(G, M)}
=\frac{\ker d^k}{\img d^{k-1}}.
\end{equation}
We emphasize that the cochain space $C^k(G, M)$, the resulting spaces $Z^k(G, M)$ and $B^k(G, M)$ all depend explicitly on the choice of the resolution $F$.
However, the resulting group-cohomology modules $H^k(G, M)$ do not depend on the choice of the resolution.
More precisely speaking, group-cohomology modules computed using different resolutions are naturally isomorphic to each other.

In the rest of this section, we give two examples to demonstrate the concept of free resolutions and their contracting homotopy.
In the first example, we show how
the inhomogeneous cocycles, which are widely used in physics literatures,
can be expressed using this language.
In fact, in math literatures, the corresponding resolution is called the bar resolution~\cite{Joyner2007}, which we shall denote by $\bar F$.
This type of resolution can be constructed for an arbitrary group $G$.
In the resulution $\bar F$, the module $\bar F_k$ is spanned by the $\mathbb ZG$ basis of the following form, $[g_1|g_2|\cdots|g_k]$, where $g_i\in G$.
The boundary operator is given as the following,
\begin{widetext}
\begin{equation}
  \label{eq:bdry-bar}
  \partial_k[g_1|\cdots|g_k]
  =g_1[g_2|\cdots|g_k]
  +\sum_{i=1}^{k-1}(-1)^i[g_1|\cdots|g_{i-1}|g_ig_{i+1}|g_{i+2}|\cdots|g_k]
  + (-1)^k[g_1|\cdots|g_{k-1}].
\end{equation}
\end{widetext}
Using this basis, a $k$-cochain $\alpha$ is represented as a function $\langle\alpha,[g_1|\cdots|g_k]\rangle$.
Rewritten as $\alpha(g_1,\ldots,g_k)$, this is the inhomogeneous cochain used in physics literatures.
Eq.~\eqref{eq:bdry-bar} gives the familiar coboundary operation of the inhomogeneous cochains,
\begin{widetext}
\begin{equation}
  \label{eq:d-bar}
  \begin{split}
  (d^k\alpha)(g_1,\ldots,g_{k+1})
  = & g_1\alpha(g_2,\ldots,g_{k+1})
  +\sum_{i=1}^k(-1)^i\alpha(g_1,\ldots,g_ig_{i+1},\ldots,g_{k+1}) \\
  &+ (-1)^{k+1}\alpha(g_1,\ldots,g_k).
\end{split}
\end{equation}
\end{widetext}

The bar resolution has the following contracting homotopy $\bar s$:
\begin{equation}
  \label{eq:s-bar}
  \bar s_k(g_0[g_1|\cdots|g_k])
  =[g_0|g_1|\cdots|g_k].
\end{equation}
We notice that, as expected, the map $\bar s_k$ does not commute with the $G$-action.
It is straightforward to check that $\bar s$ satisfies the condition in Eq.~\eqref{eq:pssp}.
Hence, it is a contracting homotopy, which confirms that $\bar F_k$ forms a long-exact sequence.
This contracting homotopy will be used in Sec.~\ref{sec:chainmap} to map inhomogeneous cochains to other basis.

The bar resolution can be cumbersome to work with, since the number of $\mathbb ZG$ basis in each module $\bar F_k$ grows exponentially with $k$, $\rank_{\mathbb ZG}\bar F_k = |G|^k$.
It is well known that one can slightly improve this by eliminating the basis elements where any one of the group element $g_i$ is $1$, the identity element of $G$.
Equivalently, in terms of inhomogeneous cocycles, one can always use coboundary equivalence to set $\alpha(g_1,\ldots,g_k)=0$ if any $g_i=1$.
The resulting resolution is called the normalized bar resolution in mathematical literatures.
In the rest of this paper, we will use $\bar F$ and $\bar s$ to denote the normalized bar resolution of a group $G$ and the associated contracting homotopy, respectively.

As an example, we examine the free resolution constructed by this algorithm for the $\mathbb Z_n$ group,
which is the chain complex of the infinite-dimensional lens space~\cite{HatcherAT}.
In this resolution, each $F_k$ is generated by only one $\mathbb ZG$-basis, denoted by $e_k$.
The boundary operator is given as the following,
\begin{equation}
\label{eq:bdry-lens}
\begin{split}
&\partial e_{2k-1} = (a-1)e_{2k-2}, \\
&\partial e_{2k} = \left(1+a+a^2+\cdots+a^{n-1}\right) e_{2k-1}.
\end{split}
\end{equation}
Here, $a$ denotes the generator of $\mathbb Z_n$ satisfying $a^n=1$.
The algorithm in HAP also constructs the following contracting homotopy of this resolution.
\begin{equation}
\label{eq:sk-lens}
\begin{split}
&s_{2k-1}\left(a^me_{2k-1}\right)=\delta_{m, n-1}e_{2k}, \\
&s_{2k}\left(a^me_{2k}\right)=
\left(1+a+\cdots+a^{m-1}\right)e_{2k+1}.
\end{split}
\end{equation}
Again, the map $s_k$ does not commute with the $G$-action.


\subsection{Chain map}
\label{sec:chainmap}

The resolution and its contracting homotopy constructed by HAP already allow us to do a wide ranges of group-cohomology calculations, including computing the classification of the group cohomology, and computing the cup and higher-cup products~\cite{brown2012cohomology,Steenrod1947,Davis1985}.
However, there are still functions of cocycles that can only be conveniently expressed using the inhomogeneous cochains~\cite{QRWangFSPT2,Morgan2018}.
The reduced resolution can still help us simplify the computation of these functions:
We first compute the cocycle functions using inhomogeneous cochains, then map the resulting inhomogeneous cocycles to the reduced resolution using a chain map, which we shall construct in this section.
In general, the chain maps between the two resolutions allow us to map cocycles between the two basis.
In the next section, we shall see that these chain maps can help us reduce the computational cost of calculating fSPT classifications.

A chain map $f$ between two resolutions $F$ and $F'$, $f: F\rightarrow F'$, is a collection of $\mathbb ZG$-linear maps $f_k:F_k\rightarrow F_k'$, such that the following diagram commutes,

\begin{equation}
\label{eq:chainmap}
\begin{tikzcd}
\cdots\arrow[r,"\partial_{k+1}"]&
F_k\arrow[r,"\partial_k"]\arrow[d,"f_k"]&
F_{k-1}\arrow[r,"\partial_{k-1}"]\arrow[d,"f_{k-1}"]&
\cdots\arrow[r,"\partial_1"]&
F_0\arrow[r,"\epsilon"]\arrow[d,"f_0"]&
\mathbb Z\arrow[d,"\id"]\\
\cdots\arrow[r,"\partial_{k+1}'"]&
F_k'\arrow[r,"\partial_k'"]&
F_{k-1}'\arrow[r,"\partial_{k-1}"]&
\cdots\arrow[r,"\partial_1'"]&
F_0'\arrow[r,"\epsilon'"]&
\mathbb Z
\end{tikzcd}
\end{equation}

Here, we describe an algorithm of constructing a chain map $f:F\rightarrow F'$ between two free $\mathbb ZG$-resolutions, using a contracting homotopy $s'$ of $F'$.
The construction is recursive.
First, at the lowest level, $f_{-1}:\mathbb Z\rightarrow\mathbb Z$ is simply the identity map.
Next, we assume that the map $f_{k-1}$ has been constructed, and proceed to construct $f_k$.
We choose a $\mathbb ZG$-basis of $F_k$, $e_{k,i}$.
Eq.~\eqref{eq:chainmap} demands that $f_k$ satisfies
\[\partial_k'f_k(e_{k,i})
=f_{k-1}(\partial_ke_{k,i}).\]
It is straightforward to check that the r.h.s is closed.
Hence, as discussed in Sec.~\ref{sec:bg}, Eq.~\eqref{eq:pssp} implies that we can choose the image of $e_{k,i}$ to be
\begin{equation}
\label{eq:fk}
f_k(e_{k,i})=s_{k-1}'f_{k-1}(\partial_ke_{k,i}).
\end{equation}
We then extend $f_k$ linearly to $F_k$.

We notice that, even with a given $s'$, the chain map $f$ constructed above is not unique. It depends on the choice of the basis in each $F_k$, because the contracting homotopy $s'$ does not commute with the $G$-action.
However, different choices of $f$ are homotopically equivalent to each other, as we shall see explicitly in Sec.~\ref{sec:cocycle-eq}.

Actually, in the above construction, only the contracting homotopy $s'$ of the second resolution $F'$ is used.
Therefore, the chain map can be constructed from an arbitrary chain complex $F$ made of free-$G$- modules, even if $F$ is not contractible.

Using a chain map $f:F\rightarrow F'$, one can map a cocycle in the basis of $F'$ to one in the basis of $F$, using the pullback map $f^\ast$.
For a cochain $\alpha^\prime \in\hhom_G(F',M)$, its image $f^\ast(\alpha)$ is given by the following relation,
\begin{equation}
\label{eq:pullback}
\forall x\in F,
\langle f^\ast(\alpha),x\rangle=
\langle\alpha,f(x)\rangle.
\end{equation}
In particular, in this work, we usually consider chain maps between two types of resolutions of $G$: $F$ is a reduced resolution given by the algorithm in HAP, and $\bar F$ is the normalized bar resolution discussed in Sec.~\ref{sec:bg}.
We denote the two chain maps between them by $f:F\rightarrow \bar F$ and $g:\bar F\rightarrow F$, respectively.
Since both $F$ and $\bar F$ have explicit contracting homotopies, both $f$ and $g$ can be constructed recursively using the algorithm in Eq.~\eqref{eq:fk}.

We end this section with an example of computing the chain maps.
Again, we consider the finite cyclic group $G=\mathbb Z_n$.
Its reduced resolution $F$, derived from the chain complex of the lens space, is given in Sec.~\ref{sec:bg}, along with a contracting homotopy.

We now demonstrate the construction of $f:F\rightarrow\bar F$.
First, since both $F_0$ and $\bar F_0$ are simply $\mathbb ZG$ with one basis, $f_0$ just maps the basis $e_0\in F_0$ to the basis $[\cdot]\in\bar F_0$.
(Recall that basis in $F_n$ are labeled by $n$ group elements. Hence, the single basis of $F_0$ is labeled by zero group element, and denoted by $[\cdot]$.)
Next, we use Eq.~\eqref{eq:fk} to construct $f_1$:
\begin{equation}
\label{eq:zn:f1}
f_1(e_1)=s_0f_0(\partial e_1)
=s_0(a[\cdot]-[\cdot])=[a].
\end{equation}
Similarly, we can proceed and compute $f_2$ and $f_3$ recursively,
\begin{align}
\label{eq:zn:f2}
f_2(e_2)&=[a|a]+[a^2|a]+\cdots+[a^{n-1}|a],\\
\label{eq:zn:f3}
f_3(e_3)&=[a|a|a]+[a|a^2|a]+\cdots
          +[a|a^{n-1}|a],\\
f_4(e_4)&=\sum_{i,j=1}^{n-1}[a^i|a|a^j|a].
\end{align}

Next, we demonstrate constructing $g:\bar F\rightarrow F$.
Comparing to $f$, the results are more lengthy.
Hence, we only compute the first two dimensions, which are used in the example of the main text.
Similar to $f_0$, $g_0$ also maps the single basis $[\cdot]$ in $\bar F_0$ to the single basis $e_0$ in $F_0$.
Next, we compute $g_1$:
\begin{equation}
  g_1([a^i]) = s_0g_0(\partial[a^i])
  =s_0(a^ie_0-e_0)=(1+\cdots+a^{i-1})e_1.
\end{equation}
In the last step, we used the contracting homotopy of the resolution in Eq.~\eqref{eq:sk-lens}.
Next, we compute $g_2$:
\begin{equation}
  \begin{split}
  g_2([a^i|a^j]) = s_1g_1(\partial[a^i|a^j])
  =s_1g_1(a^i[a^j]-[a^{i+j}]+[a^i])\\
  =s_1\left\{(1+\cdots+a^{i+j-1})e^1 - (1+\cdots+a^{l-1})e^1\right\},
  \end{split}
\end{equation}
where $l=i+j\mod n$.
Hence, if $i+j<n$, we have $l=i+j$, andthe above equation vanishes.
If $n\leq i+j<2n$, we have $l=i+j-n$, and the above equation gives
\[g_2([a^i|a^j])=s_1\left\{(a^{l-1}+\cdots+a^{l+n-1})e^1\right\}=e^2.\]
Combing these two cases, we have
\begin{equation}
  \label{eq:g2:aiaj}
  g_2([a^i|a^j])=\left\lfloor\frac{i+j}n\right\rfloor e^2,
\end{equation}
where $\lfloor(i+j)/n\rfloor$, meaning the greatest integer less than or equal to $(i+j)/n$, is 0 (1) if $i+j < (\geq) n$, respectively.

\subsection{Coboundary check}
\label{sec:coboundary}

As we discussed in Sec.~\ref{sec:motivation}, the most time-consuming task of computing an SPT classification is to check whether a obstruction function, which is a cocycle, is a trivial coboundary or not.
Such an obstruction cocycle is often expressed as an inhomogeneous cocycle.
Checking whether a cocycle is a coboundary using the normalized bar resolution is quite time-consuming, since the size of the coboundary matrix is $(|G|-1)^n$ by $(|G|-1)^{n+1}$.
In contrast, performing the coboundary check is much easier using the reduced resolution $F$, because the dimensions of the $G$ modules $F_n$ and $F_{n+1}$ are much smaller.

Hence, we propose the following approach for checking whether an inhomogeneous cocycle $\bar\alpha\in \bar Z^n(G, M)$ is a coboundary.
First, we construct a reduced resolution $F$, and the chain map $f:F\rightarrow\bar F$.
Second, we map $\bar\alpha$ to a cocycle $\alpha\in \hhom_G(F,M)$, using the pullback map, as $\alpha = f^\ast(\bar\alpha)$.
Finally, we check whether the cocycle $\alpha$ is trivial, using the reduced resolution $F$.

To be more concrete, this approach can be implemented using the following algorithm.
To check the trivialness of a $n$-cocycle $\alpha$, we use the Smith normal form of the coboundary map $d^{n-1}:\hhom_G(F_{n-1}, M)\rightarrow\hhom_G(F_n, M)$.
The Smith normal form reveals a set of invariants identifying nontrivial cocycles:
\begin{equation}
\label{eq:invariants}
I_k = \sum_ia_{k,i}\langle\alpha, e_{n,i}\rangle.
\end{equation}
A nonvanishing $I_k\neq0$ indicates that the cocycle $\alpha$ is not a coboundary.
The details of obtaining these invariants from the Smith normal form of $d^{n-1}$
are reviewed in 
Sec.~\ref{app:SNF}.
Next, we express the invariants $I_k$ with $\bar\alpha$, using the chain map $f$.
Using Eq.~\eqref{eq:pullback}, we write $\alpha(e_{n,i})$ as $\bar\alpha(f(e_{n,i}))$, and the invariants in Eq.~\eqref{eq:invariants} as
\begin{equation}
\label{eq:inv-bar}
I_k=\sum_ia_{k,i}\langle\bar\alpha,f(e_{n,i})\rangle.
\end{equation}
Finally, we compute each $I_k$ using the entries of $\bar\alpha$, and check if all $I_k$ vanish.
Any nonvanishing $I_k$ indicates that $\bar\alpha$ is a nontrivial cocycle.
Since $F$ is usually much smaller than $\bar F$ (to be more precise, the dimensions of $F_n$, $\rank_{\mathbb ZG}F_n$, are much smaller than that of $\bar F_n$),
this algorithm can save significant computational costs comparing to the naive approach using only the inhomogeneous cocycles.


We will demonstrate this algorithm and the saving on computational costs using the example of checking a 3-cycle for a cyclic group $G=\mathbb Z_n$.
As we see in Sec.~\ref{sec:bg}, the modules $F_k$ only have one $\mathbb ZG$-basis $e_k$.
The boundary operator $\partial:F_3\rightarrow F_2$ is given by $\partial e_3=(x-1)e_2$.
Hence, the corresponding coboundary operator is simply a one-by-one matrix.
If the coefficient module is $M=\uone$ with a trivial $G$-action,
the coboundary operator $d^2:\hhom_G[F_2,\uone]\rightarrow\hhom_G[F_3,\uone]$ vanishes: $d^2=0$.
Hence, there is no nontrivial coboundary equivalence, and any cocycle with a nonvanishing entry $\alpha(e_3)\neq0$ is a nontrivial cocycle.
In other words, to check the trivialness of a cocycle, we need to examine one invariant $I_1=\langle\alpha, e_3\rangle$.
Using the chain map in Eq.~\eqref{eq:zn:f3}, we express this invariant in terms of the inhomogeneous cocycle $\bar\alpha$,
\begin{equation}
\label{eq:zn:I1}
I_1=\sum_{j=1}^{n-1}\bar\alpha(a, a^j, a).
\end{equation}
This is directly related to the partition function in Eq.~\eqref{eq:Zlens}: the partition function is $Z=e^{2\pi iI_1}$.
This demonstrates that computing the invariants in Eq.~\eqref{eq:inv-bar} is equivalent to computing the partition functions on the representative $G$-bundles discussed in Sec.~\ref{sec:motivation}.

\section{Application to SPT Classification}
\label{sec:apply}

\subsection{Computing obstruction function}
\label{sec:obstruct}

We combine the algorithms introduced in Sec.~\ref{sec:algebraic} to compute the obstruction functions that appear in fSPT classification.

As an example, we discuss the obstruction function $O_5[n_2]$ in the classification of (3+1)D fSPT, which maps a Majorana decoration pattern, represented by a 2-cocycle $n_2\in H^2[G_b, \mathbb Z_2]$, to an obstruction class represented by a 5-cocycle in $H^5[G_b,\uone_T]$.
Here, we consider the simple case, where the total symmetry group $G_f$ is a direct product of the bosonic symmetry group $G_b$ and the fermion-parity symmetry $\mathbb Z_2^f$.
The more general cases where $G_f$ is a nontrivial group extension of $G_b$ over $\mathbb Z_2^f$ can be computed in a similar manner.

In terms of inhomogeneous cochains, the obstruction function is constructed in the following steps~\cite{QRWangFSPT2}:
First, one computes the $O_4[n_2]$ obstruction function, given by the following formula,
\begin{equation}
\label{eq:3d:O4}
O_4[n_2] = n_2\cup n_2.
\end{equation}
The cup product in this equation is defined in Sec.~\ref{app:cup}.
We then check whether the obstruction $O_4[n_2]$ vanishes, meaning that it is a coboundary.
This is because if it is a nontrivial cocycle, such $n_2$ will lead to violation of fermion-parity conservation and does not represent consistent Majorana-chain decorations in a 3D fSPT state.
Second, if $O_4[n_2]$ is a trivial coboundary, we need to find a solution of the equation
\begin{equation}
\label{eq:dn3=o4}
dn_3 = O_4[n_2] = n_2\cup n_2.
\end{equation}
Third, using the solution $n_3$, one can compute the obstruction $O_5$, Eq. (220) in~\cite{QRWangFSPT2}, One then needs to check if the computed $O_5$ is a trivial cocycle.
%


Although to our best knowledge, the obstruction function $O_5$ 
can only be expressed using inhomogeneous cocycle, this calculation can be accelerated using the reduced resolution and the algorithms presented in previous sections.
First, we enumerate all cohomology classes $n_2$ in $H^2[G,\mathbb Z_2]$ using cochains in the reduced resolution.
Next, we map $n_2$ to an inhomogeneous cochain, $\bar n_2 = g^\ast n_2$.
This allows us to compute $O_4[\bar n_2]$ using the cup-product formula in Sec.~\ref{app:cup}
directly.
We then check whether it is a trivial obstruction class using the algorithm in Sec.~\ref{sec:coboundary}.
(In this step, the cup product can also be computed directly in the reduced resolution, by constructing a diagonal approximation using the contracting homotopy, as described in Sec.~\ref{app:cup}.)
If $O_4[\bar n_2]$ is trivial, we can construct a solution of Eq.~\eqref{eq:dn3=o4} using the algorithm in Sec.~\ref{sec:cocycle-eq}.
We then compute $O_5$ 
and check its trivialness using the algorithm in Sec.~\ref{sec:coboundary}.

The computational cost can be further reduced using lazy evaluation, which is a commonly used method in programming and can be easily implemented in modern programming languages.
We demonstrate the use of lazy evaluation
using the example of $O_5$ and $G=\mathbb Z_n$.
Naively, to check if $O_5$ is trivial, one first computes $\bar\alpha=O_5$ 
and then check its trivialness.
Since there are $(|G|-1)^5=(n-1)^5$ entries of $\bar\alpha$,
the cost of this step scales as $(n-1)^5$.
However, using the algorithm in Sec.~\ref{sec:coboundary},
one only needs to check that all topological invariants $I_m$ vanish.
Following the steps in Sec.~\ref{sec:chainmap}, one finds that there is only one invariant, given by
\begin{equation}
\label{eq:zn:I5}
I=\sum_{1<i,j<n}\bar\alpha(a, a^i,a,a^j,a).
\end{equation}
This invariant envolves only $(n-1)^2$ entries of $\bar\alpha$.
Therefore, only these entries need to be computed from Eq.~(220) in ~\cite{QRWangFSPT2}  
Skipping the rest of the entries reduces the computational cost from $O[(n-1)^5]$ to $O[(n-1)^2]$.
In practice, one only passes the functional form of $\bar\alpha$ given by Eq.~(220) in ~\cite{QRWangFSPT2} 
instead of all its entries, to the trivialness-checking procedure.
This procedure then constructs the invariants and computes the cochain entries on the fly when they are needed.
This practice of deferring the evaluation of the entries until their values are needed is called lazy evaluation in programming.
In this way, both CPU and memory costs are saved.

\subsection{Solving cocycle equations}
\label{sec:cocycle-eq}

The reduced resolution can also be used to accelerate the task of finding one solution of the cocycle equation,
\begin{equation}
\label{eq:db=a}
d\bar\beta = \bar\alpha,
\end{equation}
where $\alpha$ is a $(k+1)$-coboundary (otherwise this equation has no solution).
This task can also be time-consuming using the inhomogeneous cocycles, as the matix form of Eq.~\eqref{eq:db=a} has dimension $(|G|-1)^k\times(|G|-1)^{k+1}$.

Naively, one may try to solve Eq.~\eqref{eq:db=a} by mapping $\bar\alpha$ to a cochain in the reduced resolution using the pullback of the chain map $f:F\rightarrow\bar F$ as $\alpha=f^\ast\bar\alpha$, find a solution $\beta=d\alpha$ there, and map it back to an inhomogeneous cocycle as $g^\ast\beta$, using $g:\bar F\rightarrow F$.
However, $g^\ast\beta$ constructed this way is not a solution of Eq.~\eqref{eq:db=a}, because the two chain maps $f$ and $g$ are not the inverse of each other.
In fact, the composition $fg$ cannot be the identity map, because the modules $\bar F_k$ have higher dimensions than $F_k$.
Instead, $fg$ is only homotopic to the identity map, meaning that it can be related to identity using a homotopy $h: fg\sim\id$.

A homotopy $h$ is a degree-1 map: $h_k: \bar F_k\rightarrow\bar F_{k+1}$, illustrated by the following diagram,
\begin{equation}
  \label{eq:hgf}
  \begin{tikzcd}
    \cdots\arrow[r,""]&\bar F_{k+1}\arrow[r,"\partial_{k+1}"]\arrow[d]&
    \bar F_k\arrow[r, "\partial_k"]
    \arrow[d, ""]
    \arrow[ld, "h_k"]&
    \bar F_{k-1}\arrow[ld, "h_{k-1}"]\arrow[d]\arrow[r]&\cdots\\
    \cdots\arrow[r,""]&\bar F_{k+1}\arrow[r,"\partial_{k+1}"]&
    \bar F_k\arrow[r, "\partial_k"]&
    \bar F_{k-1}\arrow[r]&\cdots
  \end{tikzcd}
\end{equation}
Here, the verticle arrows represent the difference between $fg$ and identity, $f_kg_k - \id_{\bar F_k}$.
The diagram is not a commutative diagram: it instead satisfies
\begin{equation}
\label{eq:h}
\partial_{k+1}h_k + h_{k-1}\partial_k
=f_kg_k-\id_{\bar F_k}.
\end{equation}
For given chain maps $f$ and $g$, the homotopy $h$ can also be constructed recursively using the contracting homotopy $s$ of $F$, in the following way similar to the algorithm in Sec.~\ref{sec:chainmap}.

Similar to the construction in Sec.~\ref{sec:chainmap},
this is done recursively.
For simplicity, we assume that both $F_0$ and $\bar F_0$ has only one $\mathbb ZG$ basis.
Therefore, $f_0$ and $g_0$ simply maps between the two unique basis, and consequently $f_0g_0$ is exactly the identity map.
As a result, we can choose $h_0=0$ because there is nothing to correct.
This is the starting point of our construction.
Next, we assume that $h_{k-1}$ has been constructed, and proceed to construct $h_k$.
We take a $\mathbb ZG$-basis $e_i$ of $\bar F_k$, and the property~\eqref{eq:h}
demands that
\begin{equation}
  \label{eq:h2}
  \partial h_k(e_i) = -h_{k-1}\partial e_i + f_kg_k(e_i)-e_i.
\end{equation}
We notice that the r.h.s. of this equation can be computed from existing constructions.
A solution of Eq.~\eqref{eq:h2} can be found using the contracting homotopy $\bar s$ of the resolution $\bar F$,
\begin{equation}
  \label{eq:hsol}
  h_k(e_i) = \bar s_k[-h_{k-1}\partial e_i + f_kg_k(e_i)-e_i].
\end{equation}
We can then extend $h_k$ linearly to $\bar F_k$.

The homotopy $h$ can be used to construct a solution of Eq.~\eqref{eq:db=a}, as it corrects the difference between $fg$ and identity.
Since it is a degree-1 map $h_k:\bar F_k\rightarrow\bar F_{k+1}$, its pullback maps a $(k+1)$-cochain $\bar\alpha$ to a $k$-cochain $h^\ast\bar\alpha$.
We can use it to augment $g^\ast\beta$ and construct a solution as
\begin{equation}
\label{eq:sln}
\bar\beta = g^\ast\beta - h^\ast\bar\alpha.
\end{equation}

We now prove that this is indeed a solution of Eq.~\eqref{eq:db=a}.
Consider any $x\in \bar F_{k+1}$.
Using the definition of the pullback maps, we get
\begin{equation}
  \langle d\bar\beta, x\rangle
  = \langle d\beta,g(x)\rangle
  - \langle\bar\alpha,h(\partial x)\rangle.
\end{equation}
Since $d\beta=\alpha$, we get
\begin{equation}
  \langle d\bar\beta, x\rangle = \langle\alpha,g(x)\rangle
  - \langle\bar\alpha,h\circ\partial(x)\rangle
  =\langle\bar\alpha,f\circ g(x)\rangle
  - \langle\bar\alpha,h\circ\partial(x)\rangle.
\end{equation}
Using Eq.~\eqref{eq:h},
we get
\begin{equation}
  \langle d\bar\beta, x\rangle =
  \langle\bar\alpha,x\rangle
  +\langle\bar\alpha,\partial\circ h(x)\rangle.
\end{equation}
Since $d\bar\alpha=0$, the second term in r.h.s. vanishes.
Hence, we conclude that $d\bar\beta = \bar\alpha$.

\section{Physical interpretation}
\label{sec:physical}

In Sec.~\ref{sec:algebraic}, we describe an algebraic algorithm that generates the topological invariants differentiating SPT phases.
The invariants generated by the algorithm coincide with the partition functions evaluated on hand-picked representative $G$-bundles.
In this section, we give an interpretation of the connection between the two.
For simplicity, we assume that $G$ is a finite unitary symmetry group, and thus consider the group cohomology with U(1) coefficients.
The results can be easily generalized to include antiunitary symmetry operations, and infinite groups.

We first review the connection between SPT states and group cohomology computed from an arbitrary resolution.
A $k$-cocycle $\alpha\in\hhom_G[F_k, \uone]$, which is a cocycle in $H^k[BG, \uone]$, can be viewed as an action, mapping each $k$-cell $\sigma\in BG_k$ to a U(1) phase factor $\langle\alpha, \sigma\rangle$.
Because $EG$ is a universal bundle, such an action can be used to construct a partition function for any $G$-bundle over a $k$-dimensional orientable space-time manifold $B$~\cite{DijkgraafWitten}.
For simplicity, we also assume $B$ is connected.
Because $G$ is a finite discrete group, the gauge connection on $B$ must vanish.
Therefore, the $G$-bundle is specified by the symmetry flux through each noncontractible loop in $B$, which can be expressed as a group homomorphism
$\gamma:\pi_1(B)\rightarrow G$.
Since $\pi_1(BG) = G$, $\gamma$ is also a homomorphism $\gamma:\pi_1(B)\rightarrow\pi_1(BG)$.
Because all higher homotopy groups of $BG$ vanish,
$\gamma$ can be further uniquely (up to homotopy) extended to a cellular map $\gamma:B\rightarrow BG$.
Algebraically, the cellular map $\gamma$ is a chain map from the chain complex of $B$ to that of $BG$, which is the resolution $F$.
Such a chain map can be constructed using the algorithm in Sec.~\ref{sec:chainmap}, using a contracting homotopy of $F$.
In particular, $\gamma$ maps each $k$-cell $\sigma\in B_k$ to an algebraic sum of $k$-cells in $BG$, denoted by $\gamma(\sigma)$.
Intuitively, this can be viewed as a decomposition of $\sigma\in B_k$ using cells in $BG$.
One can then evaluate $\alpha$ on each cell in $\gamma(\sigma)$, and define the sum of the evaluations as the value of the action on $\sigma$.
Mathematically, this is expressed as $\langle\gamma^\ast\alpha,\sigma\rangle=\langle\alpha,\gamma(\sigma)\rangle$, where $\gamma^\ast(\alpha)$ is the pullback of $\alpha$ by $\gamma$, which is a cochain on $B$, and can be viewed as an action induced by $\alpha$ defined on $B$.
Finally, one can integrate $\gamma^\ast\alpha$ on $B$, and construct the following partition function,
\begin{equation}
\label{eq:ZBAA}
Z=\exp(2\pi i\langle\gamma^\ast\alpha, [B]\rangle).
\end{equation}
Here, $[B]$ denotes the fundamental class of $B$~\cite{HatcherAT}, which is an algebraic sum of all $k$-cells of $B$, with signs given by comparing the orientation of each cell to a global orientation of $B$.

In particular, when we take $F$ to be the bar resolution, $\gamma:B\rightarrow BG$ can be viewed as a simplicial decomposition, or a triangulation, of $B$:
the image $\gamma([B])$ gives an algebraic sum of all simplices in such a decomposition,
\[\gamma([B]) = \sum s_{i_0\cdots i_k}[v_{i_0}\cdots v_{i_k}].\]
Hence, Eq.~\eqref{eq:ZBAA} becomes the following function,
\begin{equation}
  \label{eq:ZSPT0}
  Z = \exp\left\{
  2\pi i\sum_{[v_{i_0}\cdots v_{i_k}]}s_{i_0\cdots i_k}\bar\alpha(g_{i_0i_1},\ldots,g_{i_{k-1}i_k})\right\},
\end{equation}
where $g_{ij}$ is the gauge connection on the 1-cell $ij$.
For a cocycle $\bar\alpha$, the above partition function is independent of the choices of $g_{ij}$, as long as the total symmetry flux $\prod g_{ij}$ along noncontractible loops stays the same.
Therefore, one can sum over all possible choices of $g_{ij}$ and obtain the familiar form of SPT-state partition function,
\begin{equation}
  \label{eq:ZSPT}
  Z = \sum_{g_{ij}}\exp\left\{2\pi i
  \sum_{[v_{i_0}\cdots v_{i_k}]}s_{i_0\cdots i_k}\bar\alpha(g_{i_0i_1},\ldots,g_{i_{k-1}i_k})\right\}.
\end{equation}

Next, we notice that the invariants $I_l$ introduced in Sec.~\ref{sec:coboundary} can be viewed as partition functions of representative $G$-bundles.
Each bundle is based on a $k$-dimensional space-time manifold $B_l$, with a decomposition $\gamma_l:B\rightarrow BG$.
In particular, the fundamental class maps to $\gamma_l([B_l])=\sum_i a_{l,i}e_{n, i}$.
The invariant $I_l$ is then given by
\begin{equation}
  \label{eq:Il-int}
  I_l=\langle\gamma_l^\ast\alpha, [B_l]\rangle.
\end{equation}
Using Eq.~\eqref{eq:ZBAA}, we see that the partition function $Z$ is given by $\exp(I_l)$.

As an example, we revisit the representative $G$-bundle studied in Sec.~\ref{sec:motivation}.
In fact, the manifold in Fig.~\ref{fig:lens:cw} can be viewed as a CW-complex with one 3-cell, one 2-cell, one 1-cell and one 0-cell, respectively (see the caption of the figure).
Since the manifold is the three-dimensional lens space $L_3(1)$ and $BG$ is the infinite-dimensional lens space $L_3(1,1,1,\ldots)$, the chain map $\gamma:B\rightarrow BG$ maps each $k$-cell to the single $k$-cell in $BG$, which corresponds to the single generator of $F_k$ discussed in Sec.~\ref{sec:bg}.
In particular, the fundamental class of $B$ is mapped to $e_3$.
Hence, the partition function in Eq.~\eqref{eq:ZBAA} is given by $Z=e^{2\pi iI_1}$, where $I_1=\langle\alpha, e_3\rangle$.

Finally, the chain map $f:F\rightarrow\bar F$ can be viewed as a simplicial decomposition, or a triangulation, of cells in the CW-complex $BG$, whose chain complex is given by $F$.
In fact, $f$ can be viewed as a special case of celluar maps between a $G$-bundle and a classifying space of $G$, as $f:BG\rightarrow \overline{BG}$.
Here, $BG$ and $\overline{BG}$ denote a CW-complex and a simplicial complex, respectively, both serving as classifying spaces of $G$, and their chain complexes are given by $F$ and $\bar F$, respectively.
Consequently, the composition $f\circ\gamma_l:B\rightarrow\overline{BG}$ gives a triangulation of the manifold $B_l$, and the partition function
\begin{equation}
  \label{eq:ZIl}
  Z = \exp\{2\pi i\langle (f\circ\gamma_l)^\ast\alpha, [B_l]\rangle\}
  = \exp\{2\pi i\langle (f^\ast\alpha, \gamma_l([B_l])\rangle\}
\end{equation}
then computes the partition function on $B_l$ using the inhomogeneous cocycle $\bar\alpha = f^\ast\alpha$.

Combining the above understanding, we see that the invariants computed in Sec.~\ref{sec:coboundary} are the partition function of the representative $G$-bundles, computed from inhomogeneous cocycles using a triangulation.
In particular, the triangulation is constructed algebraically using the chain map $f:F\rightarrow\bar F$.
Such constructions are performed automatically by the algorithm in Sec.~\ref{sec:algebraic}.

\section{Inhomogeneous cocycles}
\label{app:inhomogeneous}

In this section, we review the inhomogeneous cocycles, a tool widely used to compute the cohomology of finite groups and to construct SPT classification.

An inhomogeneous $n$-cochain $\alpha\in C^n(G, M)$ is a function mapping $n$ group elements $g_1,\ldots,g_n$ to a coefficient $\alpha(g_1,\ldots,g_n)\in M$.
Here, $M$ is a $\mathbb ZG$ module, refered to as the coefficient module of the group cohomology.
The most common coefficient we encounter in SPT classification is the $\uone_T$ module:
In our notation, the $\uone$ module is actually a real number modulo one, which is often denoted as $\mathbb R/\mathbb Z$.
Physically, it represents a U(1) phase factor.
The subscript $T$ denotes how the symmetry group $G$ acts on this module:
$g\cdot\phi=-\phi$ if $g$ is an antiunitary operation, like the time-reversal symmetry $T$.

The coboundary operator $d^n:C^n(G, M)\rightarrow C^{n+1}(G, M)$ maps a $n$-cochain to a $(n+1)$-cochain, and it is defined as the following,

\begin{widetext}
\begin{equation}
\label{eq:app:dbdry-bar}
\begin{split}
(d\alpha)(g_1,\ldots,g_{n+1})=&
g_1\alpha(g_2,\ldots,g_{n+1})
-\alpha(g_1g_2,g_3,\ldots,g_{n+1})
+\alpha(g_1,g_2g_3,\ldots,g_{n+1})\\
&+\cdots
+(-1)^n\alpha(g_1,\ldots,g_ng_{n+1})
+(-1)^{n+1}\alpha(g_1,g_2,\ldots,g_n).
\end{split}
\end{equation}
\end{widetext}

\section{The integral group ring}
\label{app:ZG}

In this section, we briefly review the basic concepts of the integral group ring $\mathbb ZG$, and a free $\mathbb ZG$ module.

For any group $G$, we construct the integral group ring $\mathbb ZG$ as follows.
The elements of $\mathbb ZG$ are formally linear combination of group elements with integral coefficients, $x=\sum_{g\in G}x_gg$, $x_g\in \mathbb Z$.
The addition and multiplication between two elements $x=\sum_{g\in G}x_gg$ and $y=\sum_{g\in G}y_gg$ are given by
\[x+y=\sum_{g\in G}(x_g+y_g)g,\]
\[xy=\sum_{g,h\in G}x_gy_h(gh),\]
respectively.
It is straightforward to check that $\mathbb ZG$ is a ring.

For an arbitrary ring $R$, a free $R$-module $M$ can be understood as an analog of a linear space, with coefficients in $R$ instead.
The module $M$ can be generated by a $R$-basis, denoted by $e_i$.

\section{Smith Normal Form}
\label{app:SNF}

In this section, we briefly review the Smith normal form (SNF) of an integral matrix, which is used in Sec.~\ref{sec:coboundary}.

We consider an $n\times m$ matrix over $\mathbb Z$: $A=A_{ij}$.
Its SNF is a decomposition into three matrices, $L$, $R$ and $\Lambda$, such that
\begin{equation}
\label{eq:snf}
LAR = \Lambda,
\end{equation}
where $L$ and $R$ are $n\times n$ and $m\times m$ unimodular matrices, respectively, and $\Lambda$ is a diagonal matrix of dimensions $n\times m$.
As unimodular matrices, the inverse matrices of $L$ and $R$ are also integral matrices.

As an application of the SNF, we consider the coboundary condition $\alpha=d\beta$.
Here, $d$ is the coboundary map $d^{n-1}:\hhom_G[F_{n-1}, \uone]\rightarrow\hhom_G[F_n, \uone]$.
Assume that $\rank_{\mathbb ZG}F_{n-1}=m$ and $\rank_{\mathbb ZG}F_n=n$, respectively, and denote a set of $\mathbb ZG$ basis of $F_{n-1}$ and $F_n$ by $e_i^n$ and $e_i^{n-1}$, respectively.
A cochain $\alpha$ in $\hhom_G[F_n, \uone]$ is represented as a vector $\alpha_i$ using its components on $e_i^n$, $\alpha_i = \langle\alpha, e_i^n\rangle$.
Similarly, a cochain $\beta$ in $\hhom_G[F_{n-1}, \uone]$ is represented as a vector $\beta_i$ using its components on $e_i^{n-1}$, $\beta_i = \langle\beta, e_i^{n-1}\rangle$.
The coboundary map $\alpha = d^{n-1}\beta$ can then be represented as a matrix $A^i_j$, such that
\begin{equation}
  \label{eq:snf:a=Ab}
  \alpha_i = \sum_{j=1}^mA_{ij}\beta_j,
\end{equation}
According to Eq.~\eqref{eq:cobdry},
the explicit form of $A^i_j$ can be obtained by expanding $\partial e_i^n$ on basis of $e_j^{n-1}$,
\begin{equation}
  \label{eq:snf:Aij}
  \partial e_i^n = \sum_{j=1}^m A_{ij} e_j^{n-1}.
\end{equation}
Here, the group-element coefficients are converted to numbers using the group action on the coefficients.

To solve Eq.~\eqref{eq:snf:a=Ab} and check if $\alpha$ is a coboundaries, we find the SNF of matrix $A$ given by Eq.~\eqref{eq:snf}.
As a result, Eq.~\eqref{eq:snf:a=Ab} is changed into
\begin{equation}
  \label{eq:snf:La=Lb}
  L\alpha = \Lambda\beta^\prime,
\end{equation}
where $\beta^\prime = R^{-1}\beta$.
Since $R$ is unimodular, going through all possible $\beta$ is equivalent to going through all possible $\beta'$.
Hence, $\alpha$ is a coboundary if and only if there is a $\beta'$ such that Eq.~\eqref{eq:snf:La=Lb} holds.
Therefore, we consider each row in the matrix equation \eqref{eq:snf:La=Lb}, which has the following form,
\begin{equation}
\label{eq:snf:row}
\sum_{j=1}^nL_{ij}\alpha_j = \Lambda_{ii}\beta'_i.
\end{equation}
For each diagonal element $\Lambda_{ii}=0$, the LHS of Eq.~\eqref{eq:snf:row} defines an invariant, which we denote by $I^j$,
\begin{equation}
\label{eq:snf:I}
I_i = \sum_{j=1}^m L_{ij}\alpha_j.
\end{equation}
Since $\Lambda_{ii}=0$, Eq.~\eqref{eq:snf:row} implies that a coboundary must satisfy $I_i=0$.
Therefore, a nonvanishing $I_i\neq0$ indicates that $\alpha$ is not a coboundary.

\section{Cup and higher-cup products}
\label{app:cup}

In this section, we briefly review the concept of cup products and higher-cup products in group-cohomology theory, which appears frequently in formulas computing the classification of fSPT and SET phases.

The cup product is a group-cohomology operation that maps a pair of cocycles to another cocycle.
The mathematical definition of cup products can be found in Chap. V of Ref.~\cite{brown2012cohomology}.
In particular, it maps a $p$-cocycle and a $q$-cocycle to a $(p+q)$-cocycle:
\begin{equation}
  \label{eq:cup}
  \cup: H^p(G, M_1)\times H^q(G, M_2)\rightarrow
  H^{p+q}(G, M_3).
\end{equation}
Here, in general, $M_1$, $M_2$ and $M_3$ are three different $G$-modules, with a bilinear form $B:M_1\times M_2\rightarrow M_3$.

As we discuss in Sec.~\ref{sec:bg}, cocycles in a cohomology group can be expressed using different resolutions of $G$.
Using the inhomogeneous cocycles, the cup product is given by the following explicit form,

\begin{widetext}
\begin{equation}
  \label{eq:cup-inhomo}
  \alpha\cup\beta(g_1,\ldots,g_{p+q})
  =B[\alpha(g_1,\ldots,g_p),
  g_1\cdots g_p\cdot\beta(g_{p+1},\ldots g_{p+q})].
\end{equation}
\end{widetext}

In fact, this definition provides a cup product on the inhomogeneous cochains,
\begin{equation}
  \label{eq:cup-chain}
  \cup: C^p(G, M_1)\times C^q(G, M_2)\rightarrow
  C^{p+q}(G, M_3).
\end{equation}
The $\cup$ product satisfies the well-known Leibniz formula,
\begin{equation}
\label{eq:leibniz}
d(\alpha\cup\beta)
=d\alpha\cup\beta + (-1)^{\deg\alpha}\alpha\cup d\beta.
\end{equation}
This implies that if both $\alpha$ and $\beta$ are cocycles, $\alpha\cup\beta$ is also a cocycle, which is consistent with Eq.~\eqref{eq:cup}.

On an arbitrary resolution $F$ over $\mathbb ZG$, the definition of a cup product is not as straightforward as Eq.~\eqref{eq:cup-inhomo}.
In general, it requires constructing a so-called diagonal approximation $\Delta$, which is a chain map $\Delta:F\rightarrow F\otimes F$.
Here, $F\otimes F$ is the tensor product of $F$ with itself, with a diagonal $G$-action.
Using such a diagonal approximation, one can define a cup product similar to Eq.~\eqref{eq:cup-chain}:
\begin{equation}
\label{eq:cup-chain2}
\cup: \hhom_G(F_p, M_1)\times \hhom_G(F_q, M_2)
\rightarrow \hhom_G(F_{p+q}, M_3).
\end{equation}
(See Chap V of Ref.~\cite{brown2012cohomology}
for the details of the diagonal approximation and cup products.)
In particular, the cup product defined using a diagonal approximation also satisfies Eq.~\eqref{eq:leibniz}.
As a result, it also gives a cup product between cohomology classes as in Eq.~\eqref{eq:cup}.
It is important to notice that the cup product defined in Eq.~\eqref{eq:cup-chain2} is not unique, as there are many possible choices of the diagonal approximation.
However, different choices of $F$ and diagonal approximations always lead to the same cup product between cohomology classes in Eq.~\eqref{eq:cup}.

In practice, for an arbitrary resolution $R$, a cup product can be constructed in the following steps:
First, construct the tensor product $F\otimes F$ with a diagonal $G$-action.
Second, construct a chain map $\Delta: F\rightarrow F\otimes F$, which serves as a diagonal approximation, using the algorithm in Sec.~\ref{sec:chainmap}.
Last, a cup product is constructed using this diagonal approximation.
These steps can construct a cup product without the help of inhomogeneous cocycles.
Alternatively, using the ideas in Sec.~\ref{sec:obstruct},
one can compute the cup product by first mapping the cocycles to inhomogeneous cocycles, computing the cup product using Eq.~\eqref{eq:cup-inhomo}, and then mapping the result back.
In general, we expect the first approach to be more efficient, because it skips the intermediate steps involving inhomogeneous cocycles.
However, in practice, we choose to use the second approach.
This is because there are usually more complicated obstruction functions that cannot be   written entirely in terms of cup products (and higher cup products), which takes much longer to compute and can only be computed using the method in Sec.~\ref{sec:obstruct}.
Therefore, the computational cost is not a big issue here.
Consequently, we choose the second approach because it has a uniform realization with other obstruction functions.

The higher cup products can be defined in a similar way.
First, for inhomogeneous cocycles, there are explicit definitions of the higher cup products, which can be found in Ref.~\cite{Steenrod1947}.
A cup-$k$ product maps a $p$-cochain and a $q$-cochain to a $(p+q-r)$-cochain,
\begin{equation}
\label{eq:cupk-inhomo}
\cup_k: C^p(G, M_1)\times C^q(G, M_2)
\rightarrow C^{p+q-k}(G, M_3).
\end{equation}
In particular, $\cup_0$ is nothing but the cup product defined above.
Below, we give the explicit form of $\cup_1$ and $\cup_2$, which were used in obstruction functions for fSPTs.

\begin{widetext}
\begin{equation}
  \label{eq:cup1-inhomo}
  \begin{split}
  (\alpha\cup_1\beta)&(g_1,\ldots,g_{p+q-1})
  =\sum_{i = 0}^{p-1}
  (-1)^{(p-i)(q+1)}\\
  &B[\alpha(g_1,\ldots,g_i,g_{i+1}\cdots g_{i+q},g_{i+q+1},\ldots,g_{p+q-1})
  , g_1\cdots g_i\beta(g_{i+1},\ldots,g_{i+q})].
\end{split}
\end{equation}
\begin{equation}
  \label{eq:cup2-inhomo}
  \begin{split}
  (\alpha\cup_2\beta)(g_1,\ldots,g_{p+q-2})
  =\sum_{0\leq i<j\leq p}
  (-1)^{(p-i)(j-i+1)}
  B[\alpha(g_1,\ldots,g_i,g_{i+1}\cdots g_j,g_{j+1},\ldots,g_{j-i+p}),\\
  g_1\cdots g_i\beta(g_{i+1},\ldots,g_j,g_{j+1}\cdots g_{j-i+p},g_{j-i+p+1},\ldots g_{p+q-2})].
\end{split}
\end{equation}
\end{widetext}

The cup-$k$ product satisfies the following relation (Thm. 5.1 of Ref.~\cite{Steenrod1947}),
\begin{widetext}
\begin{equation}
\label{eq:leibniz-k}
d(\alpha\cup_k\beta)
=(-1)^{p+q-k}\alpha\cup_{k-1}\beta
+(-1)^{pq+p+q}\beta\cup_{k-1}\alpha
+d\alpha\cup_k\beta + (-1)^p\alpha\cup_kd\beta.
\end{equation}
\end{widetext}
As a result, the $\cup_k$ product gives a product between cohomology classes,
\begin{equation}
\label{eq:cup-k}
\cup_k:H^p(G, M_1)\times H^q(g, M_2)
\rightarrow H^{p+q-k}(G, M_3).
\end{equation}

The higher cup products can also be constructed on an arbitrary resolution, without using the inhomogeneous cocycles.
This is done using the higher diagonal approximations~\cite{Davis1985}.
The higher diagonal approximations are series of homotopy equivalences, which can be constructed recursively using the method in Sec.~\ref{sec:cocycle-eq}.
This allows us to compute higher cup products without going through the inhomogeneous cocycles.
However, in practice, we choose to use the approach of mapping to/from the inhomogeneous cocycles, for similar reasons as in the case of the cup product.

\section{The SptSet Package}
\label{sec:sptset}

The algorithm described in this work is implemented in the package SptSet for the GAP software.
It can be used to compute the classification of fSPT states protected by 2D wallpaper groups, which is listed in the main text.
Once the package is installed following the instruction on its website, the results can be computed by running the script in examples/fspt\_2d\_ez.g and examples/fspt\_2d\_s12.g, respectively.
The full design and functionality of this package will be reported elsewhere.

\bibliography{spt,cohomology}